# Towards a Model-Based Framework for Integrating Usability and Software Engineering Life Cycles

**Pardha S. Pyla, Manuel A. Pérez-Quiñones, James D. Arthur & H. Rex Hartson**

Virginia Tech, 660 McBryde Hall, Blacksburg, VA 24061, USA

{ppyla, perez, arthur, hartson}@cs.vt.edu

**Abstract:** In this position paper we propose a process model that provides a development infrastructure in which the usability engineering and software engineering life cycles co-exist in complementary roles. We describe the motivation, hurdles, rationale, arguments and implementation plan for the need, specification and the usefulness of such a model. Our approach does not merge one lifecycle's techniques into another; rather it coordinates each lifecycle's activities, timing, scope, and goals using a shared design representation between the two lifecycles. We describe potential shortcomings and conclude with an implementation agenda for this process model.

## 1 Introduction

### 1.1 Parts of interactive software systems

Interactive software systems have both functional and user interface parts. Although the separation of code into two clearly identifiable modules is not always possible, the two parts exist conceptually and each must be designed on its own terms.

The user-interface part, which often accounts for an average of half of the total lines of code (Myers and Rosson, 1992), begins as an interaction design, which finally becomes implemented in user interface software. Interaction design requires specialized usability engineering (UE) knowledge, training, and experience in topics such as human psychology, cognitive load, visual perception, task analysis, etc. The ultimate goal of UE is to create systems with measurably high usability, i.e., systems that are easy to learn, easy to use, and satisfying to their users. A practical objective is also to provide interaction design specifications that can be used to build the interactive component of a system by software engineers. In this position paper we define the usability role as that of the developer who has responsibility for building such specifications.

The functional part of a software system, sometimes called the functional core, is represented by the non-user-interface software. The design and development of this functional part requires specialized software engineering (SE) knowledge, training, and experience in topics such as algorithms, data structures, software architecture, database management, etc. The goal of SE is to create efficient and reliable systems containing the specified functionality, as well as implementing the interactive portion of the project. We define the SE role as that of the developer who has the responsibility for this goal.

To achieve the goals for both parts of an interactive system, i.e., to create an efficient and reliable system with required functionality and high usability, effective development processes are required for both UE (figure 1) and the SE lifecycles (figure 2). The UE development lifecycle is an iteration of activities for requirement analysis (e.g., needs, task, work flow, user class analysis), interaction design (e.g., usage scenarios, screen designs, information design), prototype development, and evaluation; producing a user interface interaction specification. The SE development cycle mainly consists of concept definition and requirements analysis, design (generally proceeds in two phases as shown in figure

2: preliminary and detailed design), design review, implementation, and integration & testing (I&T).

## 1.2 The problem: Coordinating the development of the two lifecycles

Given the fact that each of these development processes is now reasonably well established, that the two have the same high level goal of producing software that the user wants, and that the two must function together to create a single system, one might expect solid connections for collaboration and communication between the two development processes. However, the two disciplines are still typically separate and are applied independently with little coordination in product development. For example, it is not uncommon to find usability engineers being brought into the development process after the implementation stage. They are asked to 'fix' the usability of an already implemented system, and even then any changes proposed by the usability engineers that require architectural modifications are often ignored due to budget and time constraints. Those few changes that actually get retrofitted incur huge costs and modify the software modules that were not written anticipating changes. The lack of coordination between the usability and software engineers often leads to conflicts, gaps, miscommunication, spaghetti code due to unanticipated changes, brittle software, and other serious problems during development, producing systems falling short in both functionality and usability and in some cases completely failed projects.

In particular, there is a need within interactive system development projects for:

- communication among developer roles having different development activities, techniques, and vocabularies;
- coordinating independent development activities (usability and software engineers working together on role-specific activities);
- synchronizing dependent development activities (timely readiness of respective work products);
- identifying and realizing constraints and dependencies between the two parts.

## 1.3 Objective

The objective of our position paper is to describe a development process model that:

- integrates the two lifecycles under one common framework;
- retains the two development processes as separately identifiable processes, each with its own life cycle structure, development activities, and techniques; and
- is built upon a common overall design representation, shared by the two developer roles and processes.

The common design representation is the key to coordination of interface and functional core development, to communication among different developer roles, and identification and realization of constraints and dependencies between the two parts. The common design representation also identifies the possibility of future changes. This allows the two developer roles to

- design for change by keeping the design flexible, and to
- mitigate the changes that could be imposed on each lifecycle.

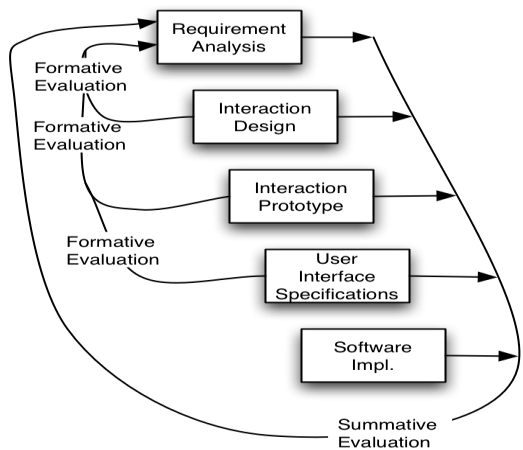

**Figure 1: Usability engineering process model**

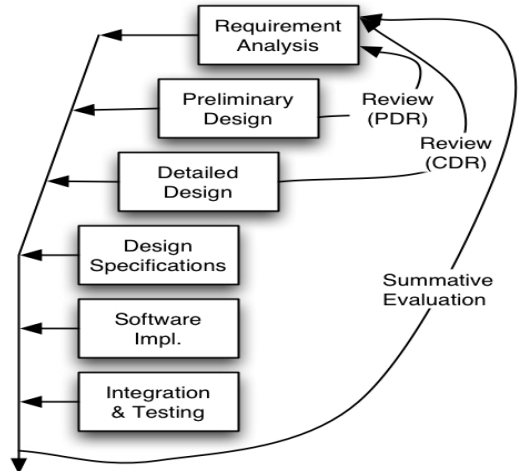

**Figure 2: Software engineering process model**

# 2 Background

## 2.1 Operating assumptions

A strong operating requirement for our work is to maintain UE and SE as separate lifecycle development processes for the two parts. It is not our goal to merge either development process into the other, but to establish a development infrastructure in which both can exist and function in parallel. UE and SE processes each require special knowledge and skills. Trying to integrate, for example, the UE lifecycle into the SE lifecycle, as done in (Ferre, 2003), creates a risk (and a high likelihood) of deciding conflicts in favor of software development needs and constraints, and against those of usability.

## 2.2 Similarities between lifecycles

At a high level, UE and SE share the same objectives:
- Seeking to understand the client's, customer's, and users' wants and needs;
- Translating these needs into system requirements; and
- Designing a system to satisfy these requirements
- Testing to help ensure their realization in the final product.

## 2.3 Differences between lifecycles

The objectives of the SE and UE are achieved by the two developer roles using different development processes and techniques. At a high level, the two lifecycles differ in the requirements and design phases but converge into one at the implementation stage (figure 3) because ultimately software developers implement the user interface specifications. At each stage, the two lifecycles have many differences in their activities, techniques, timelines, iterativeness, scope, roles, procedures, and focus. Some of the salient differences are identified here.

### 2.3.1 Different levels of iteration and evaluation

Developers of interaction designs often iterate early and frequently with design scenarios, screen sketches, paper prototypes, and low-fidelity, roughly-coded software prototypes before much, if any, software is committed to the user interface. Often this frequent and early iteration is done on a small scale and scope, often to evaluate a part of an interaction design in the context of a small number of user tasks. Usability engineers evaluate interaction designs in a number of ways, including early design walk-throughs, focus groups, usability inspections, and lab-based usability testing with the primary aim of finding errors in the design.

Software engineers identify the problem, decompose and represent the problem in the form of requirements (requirements analysis block in figure 2), transform the requirements into design specifications (preliminary and detailed design blocks in figure 2) and implement these design specifications. Traditionally these activities were performed using the rigid sequential waterfall model. Later these basic activities were incorporated into a more iterative spiral model (Boehm, 1988), with a risk analysis and an evaluation activity at the end of each stage. Even though these new development models, such as the spiral model are evolving towards the UE style by accommodating and anticipating changes at each iteration, functional software development for a system, for most part, is iterated usually on a larger scale and scope. The testing in this SE lifecycle is primarily done at the end and for verifying the implementation of the system and aims at checking the compliance of the system to specifications, completeness, and to ensure the accuracy of the integration stage.

### 2.3.2 Differences in terminology

Even though certain terms in both lifecycles sound similar they often mean different things. For example:
- in UE, 'testing' is a part of design and primarily aims at validating the design decisions (identified as formative evaluation in figure 1) whereas in SE 'testing' is an independent stage with the primary aim to check the implementation of the system and to verify its conformance to specifications. Analysis and validation of the design specifications performed in SE is often called 'review' (identified in figure 2) and once the specifications pass the review stage, they become a binding document between the client and the development team.
- a (use case) scenario in SE is used to "identify a thread of usage for the system to be constructed (and) provide a description of how the system will be used" (Pressman, 2001). Whereas in UE, a scenario is "a narrative or story that describes the activities of one or more persons, including information about goals, expectations, actions, and reactions (of persons)" (Rosson and Carroll, 2002).
- the SE group refers to the term 'develop' to mean creating software code, whereas the usability engineers use 'develop' to mean iterate, refine, and improve usability

Overall, the software engineers concentrate on the system whereas the usability engineers concentrate on users. Such fundamental difference in focus is one more reason why it is difficult to merge these two lifecycles.

*2.3.3 Differences in requirements representation*

Most requirement specifications documented by software engineers use plain English language and are generally very detailed. These specifications are written specifically to drive the SE development process. On the other hand, usability engineers specify interactive component issues such as feedback, screen layout, colors, etc. using artifacts such as prototypes, use cases, and screen sketches. These artifacts are not detailed enough to derive software specifications, instead they require additional refinement and design formulation before implementation. Therefore, they cannot be used to drive the software development process directly.

# 3 Current Practices

In spite of the extensive research and maturity achieved in each of these lifecycle areas, there has been a marked deficiency of understanding between the two. In general, the two teams do not understand the others' goals and needs and do not have an appreciation for the other's area of expertise. One apparent reason for this situation is the way computer science courses are typically offered in colleges: SE courses often omit any references to user interface development techniques (Douglas et al., 2002) and UE courses do not discuss the SE implications of usability patterns.

## 3.1 Lack of coordination

When translated into development activities, this lack of understanding between the two developer roles often leads to working separately as shown in figure 3, when they could be more efficient and effective working together. For example, both roles must do some kind of field studies to learn about client, customer, and users wants and needs, but they often do this without coordination. Software engineers visit the customers for functional requirements elicitation (Pressman, 2001), determination of physical properties and operational environments of the system (Lewis, 1992), etc. Usability engineers visit clients and users to determine, often through "ethnographic studies", how users work and what they need for computer-based support for that work. They seek task information, usage scenarios, and user class definitions. Why not do this early systems analysis together? Much value can be derived from working together on system analysis and requirements gathering in terms of team building, communication, and each lifecycle expert recognizing the value, and problems, of the other, and early agreement on goals and requirements. Instead, each development group reports its results in documentation not usually seen by people in the other lifecycle; each just uses the results to drive their part of the system design and finally merge at the implementation stage (figure 3).

Another important shortcoming of the practice shown in figure 3 is the fact that the independently generated user interface specifications on UE side and the design specifications on SE side are submitted to the development team at implementation stage. It is however, critical, to have the user interface specifications before a design document can be generated by the software engineers because of the many dependencies and constraints between the interface on the functional core and vice versa. Moreover, such lack of coordination of development activities presents a disjointed appearance of the development team to the client. It is likely to cause confusion on the clients: "why are we being asked similar questions by two different groups from the same development team?"

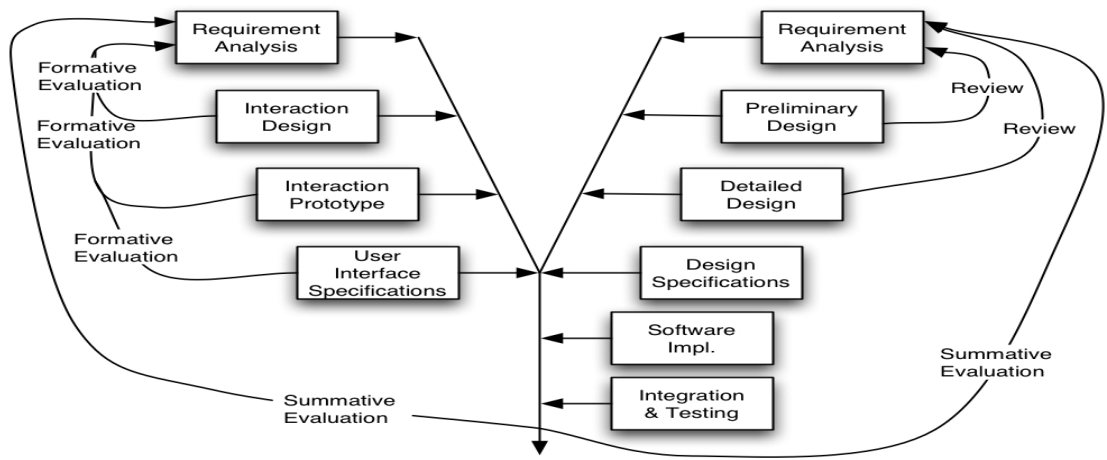

**Figure 3: Current practices: Processes without communication/coordination**

### 3.2 Lack of provision for change

In interactive systems development, every iteration brings change. This change often affects both lifecycles because of the various dependencies that exist between the two process models. One of the most important requirements for system development is to identify the possibility for change and to design accordingly. Another important requirement is to try to mitigate the extent of change by coordinating the activities of the two lifecycles and by agreeing on a common structure upon which each developer role can base their design. The more the two developer roles work without a common structure (figure 3); the greater the possibility that the two designs efforts will have major incompatibilities.

### 3.3 Lack of synchronization of development schedules

Even though developers from each process can do much work in parallel, there are times when both the lifecycles must come together for various checkpoints. These checkpoints must be verified by the process during a verification and the validation stage. For example, during a final or near-final UE testing phase (identified as the formative evaluation in figure 3), both the usability and the software engineers should be present to integrate their parts and refine test plans.

However, as shown in figure 3, the more each team works independently of the other, the less likely both groups will be scheduling its development activities to be ready for the common checkpoints.

### 3.4 Lack of communication among different developer roles

Although the roles can successfully do much of their development independently and in parallel, a successful project demands that the two roles communicate so that each knows generally what the other is doing and how that might affect its own activities. Each group needs to know how the other group's design is progressing, what development activity they are currently performing, what insights and concerns they have about the project, and so on. But the current practice (figure 3) does permit such communication to take place because the two lifecycles operate independently and there is no structured process model to facilitate the communication between these two lifecycles.

### 3.5 Lack of constraint mapping and dependency checks

Because each part of an interactive system must operate with the other, many system requirements have both a user interface and a functional part. When the two roles gather requirements separately and without communication, it is easy to capture the requirements that are conflicting and incompatible. Even if there is some form of communication between the two groups, it is inevitable that some parts of the requirements or design will be forgotten or will "fall through the cracks."

As an example, software engineers perform a detailed functional analysis from the requirements of the system to be built. Usability engineers perform a hierarchical task analysis, with usage scenarios to guide design for each task, based on their requirements. Documentation of these requirements and designs is kept separately and not necessarily shared. However, each view of the requirements and design has elements that reflect counterpart elements in the other view. For example, each task in the task analysis can imply the need for corresponding functions in the SE specifications. Similarly, each function in the software design can reflect the need for support in one or more user tasks in the user interface. When some tasks are missing in the user interface or some functions are missing in the software, the respective sets of documentation are inconsistent, a detriment to success of the project.

Sometimes the design choices made in one lifecycle constrain the design options in the other. For example, in a PDA based navigational application, a usability feature could be that the application core automatically determines the location and provides navigational assistance accordingly. But for this feature to materialize the backend core should have the capability and processing power to use a GPS to pinpoint the user's location. This is an example of the backend core limiting the interaction (usability) feature. Similarly, a data visualization tool that uses dynamic queries (Ahlberg and Wistrand, 1995), in which a user is able to move a slider on the user interface to change (filter) an attribute in the database being visualized is an example of interaction feature limiting backend core. This tool requires that the user interface be refreshed at the speed of the user action on the slider. This puts a finite time limit on the number of records in the database that the backend core can query and refresh the visualization within this time interval.

The intricacies and dependencies between user interface requirements and backend functionality have begun to appear in the literature. For example, in (Bass and John, 2001), user interface requirements and styles, such as support for undo, are mapped to particular software architectures required for the implementation of such features.

Because of the constraints on one another, independent application of the two life cycles (figure 3) would almost certainly fail and an integrated process model that facilitates communication between these two lifecycles is very essential.

## 4 Proposed Process Model

### 4.1 Activity awareness and lifecycle independence

Using our process model (figure 4), each developer role can see their own and the other's lifecycle status, activities, the iteration of activities, the timeline, techniques employed or yet to be employed, the artifacts generated or yet to be generated, and the mappings between the two domains if present. The view of each role would show only those activities that are relevant to that role. Each role views the shared design representation through its own filters (figure 5) so that, for example, the software engineers see only the software implications that result from the previously mentioned iterativeness in UE, but not the techniques used or the procedure followed. Similarly, if software engineers need iteration to try out different algorithms for functionality, it would not affect the usability lifecycle. Therefore, the process of iteration is shielded from the other role, only functionality changes are viewable through the UE filter. Each role can contribute to its own part of the lifecycle and the model allows each role to see a single set of design results, but through its own filter. Our process model places these connections and communication more on product design and less on development activities. This type of 'filter' acts as a layer of insulation, between the two processes, i.e. the process model helps isolate the parts of the development processes for one role that are not a concern of the other role. The layer needs to be concrete enough to serve the purposes, but not over specified so as to restrict the software design that will implement the user interface functionality. This prevents debates and needless concerns comparing processes and distrust on the other's techniques. Because our process model does not merge, but integrates, the two development processes, experts from one lifecycle need not know the language, terminology, and techniques of the other, and therefore can function independently.

### 4.2 User interface and functional core communication layer

Our process model advocates the need for the two development roles to specify a common communication layer between the user interface and the functional core parts. This layer is similar the specification of the communication between the model and the other two parts (view and controller) in the model view controller architecture (Krasner and Pope, 1988). This communication layer describes the semantics and the constraints of each lifecycle's parts. For example, the usability engineer can specify that an undo operation should be supported at a particular part of the user interface and that in the event of an undo operation being invoked by the user, a predetermined set of actions must be performed by the functional core. This type of communication layer specification, which will be recorded by our process model, allows the software engineers to proceed with the design by choosing a software architecture that supports the undo operation (Bass and John, 2001). How the undo operation is shown on the user interface does not affect the SE activities. This type of early specification of a common communication layer by the two lifecycles minimizes the possibility of change on the two lifecycle activities. However, this common communication layer specification might change with every iteration and these changes should be made taking into account the implications such a change will have on the already completed activities and the ones planned for the future.

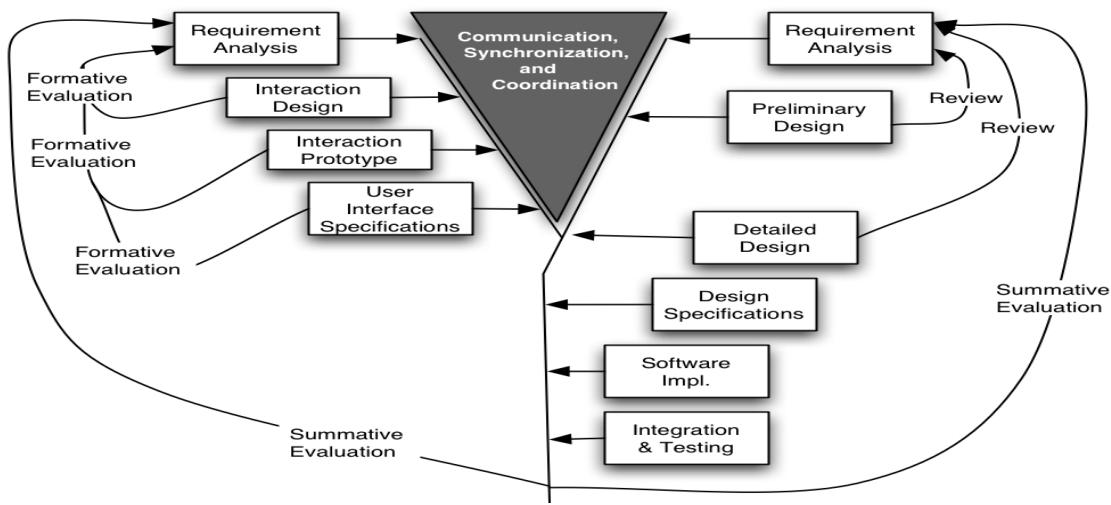

**Figure 4: Proposed model: Processes with communication/coordination**

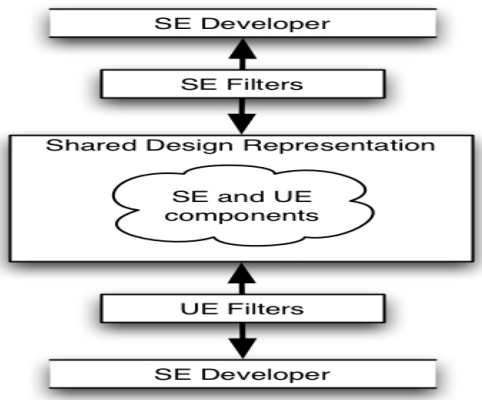

**Figure 5: Shared design representation**

### 4.3 Coordination of life cycle activities

Our process model coordinates schedules and specifies the various activities that have commonalities within the two lifecycle processes. For such activities, the process model indicates where and when those activities should be performed, who the involved stakeholders are, and communicates this information to the two groups. For example, if the schedule says it is time for usability engineers to visit the clients/users for ethnographic analysis, the process model automatically alerts the software engineers and prompts them to consider joining the usability team and to coordinate for the SE's user related activities such as requirements analysis, etc.

### 4.4 Communication between development roles

Another important contribution of this process model is the facilitation of communication between the two roles. Communication between the two roles takes place at different levels during the development lifecycle. The three main levels in any development effort are: requirements analysis, architecture analysis, and design analysis. Each of these stages results in a set of different artifacts based on the lifecycle. The process model has the functionality to communicate these requirements between the two domains. For example, at the end of UE task analysis the usability group enters the task specifications into the model and the SE group can view these specifications to guide their functional decomposition activities. At the end of such an activity, the SE group enters their functional specifications to the model for the usability people to cross check. This communication also helps in minimizing the effects of change and the costs to fix these changes. By communicating the documents at the end of each stage, the potential for identifying errors or incompatibilities increases as compared to waiting till the usability specifications stage. This early detection of mismatches is important because the cost to fix an error in the requirements that is detected in the requirements stage itself is typically four times less than fixing it in the integration phase and 100 times less than fixing it in the maintenance stage (Boehm, 1981).

### 4.5 Constraints and dependencies

The design representation model incorporates automatic mapping features, which will map the SE and UE part of the overall design based on their dependencies on each other. For example, there exists a many-to-many mapping between the tasks on the user interface side and the functions on the functional side. In the event of identifying a new task after a particular iteration by the usability group, the design representation model will automatically alert the software group about the missing function(s) and vise versa. So when the software engineer tries to view the latest task addition s/he is given a description that clearly describes what the task does and what the function should do to make that task possible. This way the developers can check the dependencies at regular time intervals to see that all the tasks have functions and vice versa and that there are no 'dangling' tasks or functions that turn up as surprises when the two roles finally do get together.

### 4.6 Potential downsides of the model

Our process model has the following downsides due to the various overheads and additional tasks that arise because of the coordination of the two lifecycles:

- Increase in the overall software development lifecycle;
- Additional effort required by the experts in each lifecycle for document creation and entry into the design representation model;
- Additional effort required for coordination of various activities and schedules;
- Need for stricter verification process than conventional processes to enforce the various synchronisation checkpoints during the development effort; and
- Resource overhead to carry out all the above mentioned drawbacks.

## 5 Implementation and Test Plan

In order to build the above described integrated process model we plan to use the following agenda:

- Analyze each lifecycle in maximum detail and catalogue all activities and widely accepted domain terminology.
- Investigate each of the activities in both domains and list the artifacts that could result from these activities. These artifacts would become a part of the 'view' for the other group.
- Determine potential overlaps in these activities within the two processes and specify the type of overlap. Based on the type of overlap, the role of the developer who can perform this activity will be identified. This information would be used to alert the groups of a possible overlapping activity and also provide the profile of the person suitable for that activity.
- Investigate terminology mismatches in each of the overlapped activities and prepare a handbook for the experts.
- Create a list of potential dependencies or constraints between parts of software system design as produced by the two roles using the literature available and research any missing ones. An example of such a work is the use of architectural patterns to support usability (Bass and John 2001).
- Test the framework using a project in simulated real life settings. We plan to do this by offering the SE and UE courses in an academic semester and having half the teams use the current practices and the other half use our framework.